%% file: main.tex
\documentclass{INTERSPEECH2023}

\interspeechcameraready

\input{00.commands}
\input{00.title}

\begin{document}
\maketitle

\input{01.abstract}
\input{02.introduction}
\input{03.relatedworks}
 
\input{04.method}
\input{05.results}

\input{07.conclusion}
\input{09.supp}
\bibliographystyle{IEEEtran}
\bibliography{main}
\end{document}

%% file: 00.commands.tex
\usepackage{times}
\usepackage{epsfig}
\usepackage{graphicx}
\usepackage{amsmath}
\usepackage{amssymb}
\usepackage{color}

\usepackage{microtype}
\usepackage{adjustbox}
\usepackage{listings}
\usepackage{pythonhighlight}
\usepackage{float}
\usepackage{tablefootnote}
\usepackage{tabularx}
\usepackage{booktabs}
\usepackage{float}
\usepackage{wrapfig}

\newcommand{\etal}{\textit{et al}.}

\newcommand{\rom}[1]{\uppercase\expandafter{\romannumeral #1\relax}}

\usepackage{pifont}
\newcommand{\cmark}{\ding{51}}%
\newcommand{\xmark}{\ding{55}}%

%% file: 00.title.tex
\title{Towards Attention-based Contrastive Learning for Audio Spoof Detection}
\name{Chirag Goel, Surya Koppisetti, Ben Colman, Ali Shahriyari, Gaurav Bharaj}
\address{Reality Defender Inc., USA}
\email{\{chirag,surya,ben,ali,gaurav\}@realitydefender.com}


%% file: 01.abstract.tex
\begin{abstract}
Vision transformers (ViT) have made substantial progress for classification tasks in computer vision. Recently, Gong et. al. '21, introduced attention-based modeling for several audio tasks. However, relatively unexplored is the use of a ViT for audio spoof detection task. We bridge this gap and introduce ViTs for this task. A vanilla baseline built on fine-tuning the SSAST (Gong et. al. '22) audio ViT model achieves sub-optimal equal error rates (EERs). To improve performance, we propose a novel attention-based contrastive learning framework (SSAST-CL) that uses cross-attention to aid the representation learning. Experiments show that our framework successfully disentangles the bonafide and spoof classes and helps learn better classifiers for the task. With appropriate data augmentations policy, a model trained on our framework achieves competitive performance on the ASVSpoof 2021 challenge. We provide comparisons and ablation studies to justify our claim.

\end{abstract}

%% file: 02.introduction.tex
\section{Introduction}
\label{sec:introduction}

Vision Transformers (ViTs) (Dosovitskiy~\etal~\cite{Dosovitskiy2020AnII}) have emerged as the state-of-the-art method for multiple computer vision~\cite{Hatamizadeh2022GlobalCV} and audio (Gong~\etal~\cite{gong2022ssast}) tasks. Audio ViTs can \textit{self-attend} to extract optimal patch embeddings and are capable of learning long-range global context within spectrograms \cite{Gong2021ASTAS, gong2022ssast}. Inspired by the success of audio ViTs in sound and speech classification, here we introduce ViTs for the audio spoof detection task.

Similar to \cite{gong2022ssast}, the standard approach is to fine-tune a pretrained audio ViT for a given downstream classification task using cross-entropy. Pre-training is done on a large dataset, such as AudioSet~\cite{Gemmeke2017AudioSA} or LibriSpeech~\cite{Panayotov2015LibrispeechAA}, and fine-tuning on a smaller task-specific dataset. However, this standard approach does not empirically work well for audio spoof detection (see Section~\ref{sec:contrastive_impact}), and results in high equal error rate (EER). Pretraining on a large audio dataset such as AudioSet-2M~\cite{Gemmeke2017AudioSA} is a key requirement for audio ViTs to perform well~\cite{Gong2021ASTAS, gong2022ssast}. However, such \textit{bonafide-only} datasets are not optimal as they don't contain spoof samples; spoof detection is an out-of-distribution downstream task.

 The \textit{de-facto} dataset for audio spoof detection is ASVSpoof 2021 logical access (LA) challenge~\cite{yamagishi2021asvspoof}. Its training set is the same as ASVSpoof2019~\cite{Nautsch2021ASVspoof2S} and contains clean studio quality data. Its test set, on the other hand, contains data corrupted by codecs and transmission channel artifacts. Also, when finetuning an audio ViT for spoof detection, the ASVSpoof19 LA training dataset falls under the small data regime.\footnote{Contains $25,380$ samples, which is about $1\%$ of AudioSet-2M~\cite{Gemmeke2017AudioSA}.} We thus require an appropriate training and augmentation strategy that helps achieve robustness against data corruption in the test set.

Supervised contrastive learning (CL) methods ~\cite{Khosla2020SupervisedCL} that use Siamese networks (Koch~\etal~\cite{koch2015siamese}) have emerged as an algorithm class that helps learn efficient data representations. Such approaches learn a representation space from generic input features by pulling samples from the same class closer together and pushing samples from different classes apart, even on limited data. Thus, they are able to train on small datasets, learn better classification margins, and are more robust to data corruptions than cross-entropy classifiers \cite{Khosla2020SupervisedCL}. Inspired by these merits, we pursue a CL approach. We introduce a cross-attention branch into the training framework and propose a novel loss formulation that measures the (dis-)similarity between the self and cross-attention representations to separate the bonafide and spoof classes.

We observe that the proposed attention-based CL framework, with appropriate data augmentations, is able to learn \textit{discriminative representations} that disentangle the bonafide samples from the spoofed ones. Further, our approach shows a significant gain in performance over the baseline cross-entropy classifier. To summarize, the main contributions of our work are:
\begin{enumerate}
\item We propose a two-stage contrastive learning framework to train an audio ViT for the spoof detection task.

\item  We consider Siamese training for representation learning and introduce a cross-attention branch into the training framework to learn discriminative representations for bonafide and spoof classes;  a suitable loss function is formulated.

\item A MLP classifier trained on the learned representations outperforms the ASVSpoof 2021 challenge baselines~\cite{yamagishi2021asvspoof} and competes with the best-performing models.
\end{enumerate}

Finally, while in this paper we focus on audio spoof detection, the proposed CL framework is not constrained by design to our specific problem, and is a general-purpose framework. In the future, we plan to explore the framework for other downstream tasks, such as language identification and acoustic scene classification.

%% file: 03.relatedworks.tex
\section{Related works}
\label{sec:rw}

\begin{figure*}[t]
  \centering
\includegraphics[width=\linewidth] {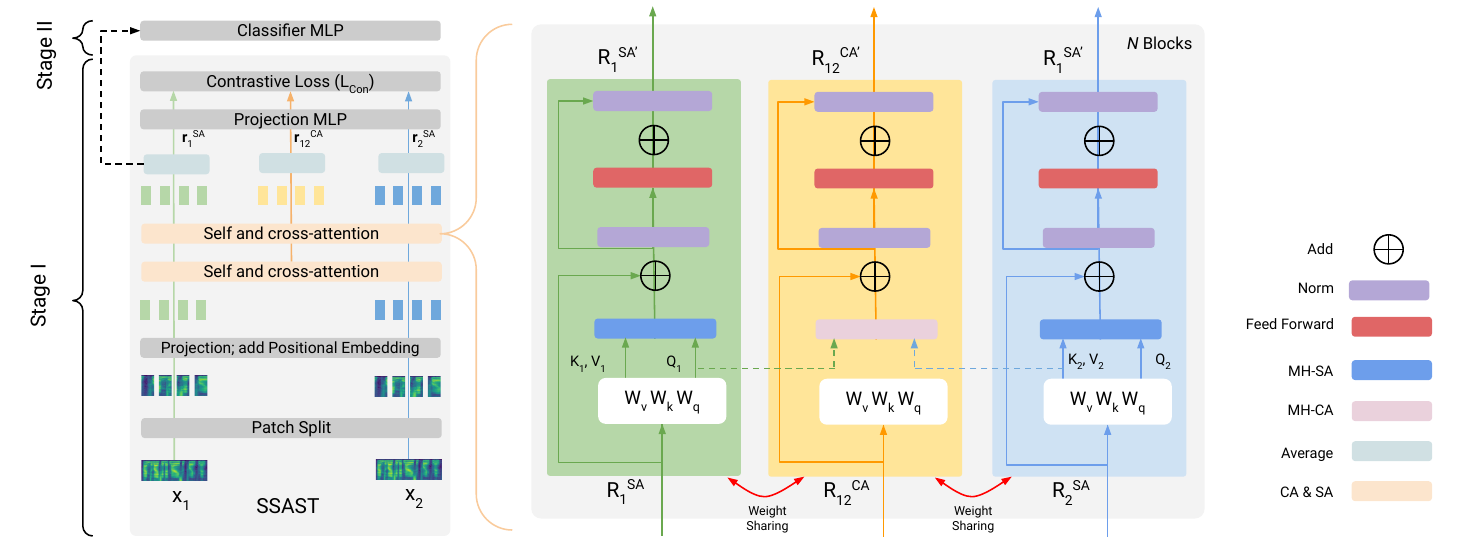}
    \caption{SSAST-CL: A two-stage contrastive learning framework to train the SSAST model \cite{gong2022ssast} for audio spoof detection. In Stage \rom{1}, we employ Siamese training with weight-sharing across two multi-head self-attention (MH-SA) and one multi-head cross-attention (MH-CA) branches. Model weights are learned using a contrastive loss which measures the (dis-)similarity between the self and cross-attention representations $(\mathbf{r}_1^{SA}, \mathbf{r}_2^{SA}, \mathbf{r}_{12}^{CA})$. In Stage II, a MLP classifies the learned representations as bonafide or spoof. }
    \label{fig:training_framework}
    \vspace{-10px}
\end{figure*}

\textbf{Attention-based modeling.} Self-attention mechanism relates different patches within an input image using attention weights and encodes efficient representation of an image. It is a building block in ViTs, and has facilitated state-of-the-art (SOTA) classification performance on multiple computer vision tasks \cite{Dosovitskiy2020AnII,Qian2022EntroformerAT,Yuan2021VOLOVO}. Cross-attention~\cite{Chen2021CrossViTCM,Xu2021CDTransCT} extends the idea to image pairs and uses attention matrices to encode an aggregate representation for the input pair based on the similarity in their patches. Due to its ability to align features from different inputs, cross-attention has been used for domain adaptation \cite{Xu2021CDTransCT} and multi-modal feature fusion \cite{Li2018NeuralSS,Hu2021UniTMM}. ViTs with self-attention only \cite{gong2022ssast} have achieved SOTA performance on multiple audio/speech classification tasks, however use of both self and cross-attention in audio ViTs, especially for audio spoof detection, is unexplored.

\noindent \textbf{Contrastive learning for audio classification.}
In CL, samples are fed into the network in form of data pairs in order to contrast between them when learning the representation spaces of each class. A typical CL formulation maximizes the similarity of samples in an input pair when they belong to the same class, and their dissimilarity otherwise~\cite{saeed2021contrastive, Xie2021SiameseNW}.  In \cite{saeed2021contrastive, Xie2021SiameseNW}, CL was used to derive audio representations for a variety of speech and non-speech classification tasks, including speaker identification, acoustic scene classification, and music recognition.

Xie~\etal~\cite{Xie2021SiameseNW} focus on detecting studio quality spoofed audio and proposed a CL algorithm where a Siamese CNN is trained on the ASVSpoof19 LA. On the other hand, we propose a CL formulation that uses cross-attention to train a ViT for spoof detection. Further, it is unclear whether a Siamese network can perform well on the ASVSpoof 2021 LA, which contains impairments from transmission codecs~\cite{yamagishi2021asvspoof}. To this end, we use appropriate data augmentations that help achieve robustness against such codec impairments.

\noindent \textbf{Spoof detection.} Early works have focused on handcrafted features, such as cepstral coefficients~\cite{Sahidullah2015ACO}, with classical methods such as Gaussian mixture models \cite{Todisco2017ConstantQC} (also see \cite{Nautsch2021ASVspoof2S} and references therein). Recent methods use CNNs that operate on 2D features extracted from audio signals~\cite{gao2021generalized, zeghidour2021leaf, Lavrentyeva2019STCAS}, or directly map raw waveforms to their labels in an end-to-end fashion~\cite{Tak2020EndtoEndAW, hua2021towards}. Introduction of attention mechanism into CNN models has further enhanced the single model performance for the ASVSpoof21 challenge~\cite{Liu2022ASVspoof2T}, particularly those employing temporal and/or spatial attention. For this challenge, the works \cite{tak2022automatic} and \cite{eom2022anti} employ wav2vec2.0~\cite{baevski2020wav2vec} large audio model as a feature extractor and achieve significantly better EER. Among other contributors, the improvement comes from wav2vec2.0 - a deep network of CNN and transformer blocks. 
While it is clear from SOTA that the introduction of attention layers in CNNs gave improved performance, it is unclear whether a pure attention-based model can be used for spoof detection. We bridge this gap and study the use of ViTs for audio spoof detection.

%% file: 04.method.tex
\section{Method}
\label{sec:method}

 \begin{figure*}[h]
  \centering
\includegraphics[width=\linewidth]{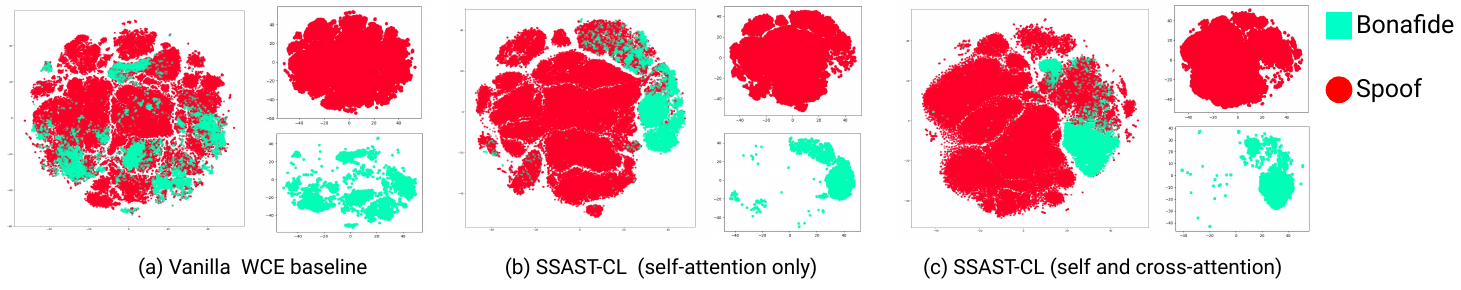}
    \caption{t-SNE embeddings for the (a) vanilla WCE baseline \cite{gong2022ssast} and (b-c) the proposed SSAST-CL solution on ASVSpoof 2021 dataset.}
    \label{fig:tsne}
    \vspace{-10px}
\end{figure*}

We introduce an audio ViT to learn efficient representations for the spoof detection task. We consider the self-supervised audio spectrogram transformer (SSAST) \cite{gong2022ssast}, a ViT pretrained on the AudioSet~\cite{Gemmeke2017AudioSA} and LibriSpeech~\cite{Panayotov2015LibrispeechAA} datasets. We propose a CL-based two-stage training framework, summarized in Fig. \ref{fig:training_framework}. In Stage I, we adapt the SSAST to train in a Siamese manner (we call it SSAST-CL), in order to learn discriminative representations for the bonafide and spoof classes. In Stage II, we train an MLP using the SSAST-CL backbone to classify the learned representations as bonafide or spoof.

\subsection{Stage I - representation learning} 

In Stage I, the goal is to learn discriminative representations for the bonafide and spoof classes. To this end, we propose the SSAST-CL framework which adapts the SSAST architecture in \cite{gong2022ssast} to a three-branch Siamese training (see Fig.~\ref{fig:training_framework}). The framework takes a pair of data $(x_1, x_2)$ as input and feeds them to SSAST-CL, whose weights are shared across the three branches. Two branches in the SSAST-CL use self-attention to compute representations $\mathbf{r}_1^{SA}$ and $ \mathbf{r}_2^{SA}$, whereas the third branch uses cross-attention to compute an aggregate representation $\mathbf{r}_{12}^{CA}$ for the input pair. 

In the self-attention branch for $x_1$ (and similarly for $x_2$), the transformer blocks encode the intermediate representations $R_1^{SA'}$ using the query, key and value matrices $(Q_1, (K_1, V_1))$ from the same $x_1$ branch. Whereas in the cross-attention branch, the transformer blocks encode the intermediate representations $R_{12}^{CA'}$ using the query $Q_1$ from the $x_1$ branch and key-value matrices $(K_2, V_2)$ from the $x_2$ branch (as shown in Fig. \ref{fig:training_framework}). By way of this design, the final representation $\bold{r}_{12}^{CA}$ becomes an aggregate representation of the input pair because it captures the information in {$x_2$ that is relevant to $x_1$.}

When computing the attention matrices in the transformer blocks, the query, key, and value are assigned learnable weights $\mathbf{W}_q$,   $\mathbf{W}_k$, and $\mathbf{W}_v$ respectively. These weights are shared across the three branches, as is the case with all other learnable parameters in SSAST-CL. Finally, we add a linear projection MLP at the end of SSAST-CL to upsample all the attention-based representations $(\mathbf{r}_1^{SA}, \mathbf{r}_2^{SA}, \mathbf{r}_{12}^{CA})$ for the contrastive loss calculations.

The model weights are optimized using a novel contrastive loss formulation which measures the (dis-)similarity between the self and cross-attention representations to separate the bonafide and spoof classes. Similar to \cite{10.1145/3394171.3413532}, we use the {cosine distance metric}, given by $\cos(\mathbf{a},\mathbf{b}) = {(\mathbf{a}^\intercal \mathbf{b})}/{\max (||\mathbf{a}|| \cdot ||\mathbf{b}||, \epsilon)}, \epsilon > 0$, to measure the similarity between the representations\footnote{Similarity is measured after feeding the representations $\mathbf{r}_{1}^{SA}, \mathbf{r}_{2}^{SA}, \mathbf{r}_{12}^{CA}$ through the projection MLP (see Fig. \ref{fig:training_framework})}. We then define the contrastive loss $L_{con}$ as:

\begin{align}
         & L_{con}  = L_{SA} + \alpha L_{CA}, \text{where }, \nonumber \\
         & L_{SA} = -1 \begin{cases}
     \log(\cos(\mathbf{r}_{1}^{SA},\mathbf{r}_{2}^{SA})) ,& c(x_1)=c(x_2) \\
     \log(1-\cos(\mathbf{r}_{1}^{SA},\mathbf{r}_{2}^{SA})) ,&  c(x_1)\neq c(x_2) \end{cases}, \nonumber \\
      & L_{CA}  = -1 
      \begin{cases}
     \log(\cos(\mathbf{r}_{1}^{SA},\mathbf{r}_{12}^{CA})) ,& c(x_1)=c(x_2) \\
     \log(1-\cos(\mathbf{r}_{1}^{SA},\mathbf{r}_{12}^{CA})) ,&  c(x_1)\neq c(x_2)
     \end{cases}
     \label{eq:loss_formulation}
\end{align}

\noindent In Eq~(\ref{eq:loss_formulation}), $c(x_1)$ and $c(x_2)$ are the (bonafide or spoof) classes to which $x_1$ and $x_2$ belong, $L_{SA}$ denotes the self-attention loss, $L_{CA}$ denotes the cross-attention loss, and $\alpha \in [0,1]$ a weighting parameter. The self-attention loss $L_{SA}$ operates on representations $\mathbf{r}_1^{SA}$ and $\mathbf{r}_2^{SA}$ from the self-attention branches. It maximizes the similarity between $\mathbf{r}_1^{SA}$ and $\mathbf{r}_2^{SA}$ if the input pair $(x_1, x_2)$ belong to the same class, and their dissimilarity otherwise. The cross-attention loss $L_{CA}$ operates on the self and cross-attention representation pair $(\mathbf{r}_{1}^{SA}, \mathbf{r}_{12}^{CA})$, as they have the same query $Q_1$ for each transformer block. $L_{CA}$ maximizes the similarity between $\mathbf{r}_1^{SA}$ and its cross-attention counterpart $\mathbf{r}_{12}^{CA}$ if the input pair $(x_1, x_2)$ belongs to the same class, and their dissimilarity otherwise.  While the self-attention loss term directs the model to learn representations that separate the sample classes, the cross-attention loss term serves as a regularizer by pushing the class-specific representations away from the aggregate representations computed by the cross-attention branch. 

A traditional Siamese training network only computes the self-attention representations $\mathbf{r}_1^{SA}$ and $\mathbf{r}_2^{SA}$ and measures their (dis-)similarity to separate the bonafide and spoof classes. Here, we introduce an additional cross-attention branch into the training framework, in order to help the model learn more discriminative representations. After the training is complete, the projection MLP at the end of Stage \rom{1} is discarded, as is commonplace in contrastive learning methods \cite{chen2020simple}. 

\subsection{Stage II - classifier \label{sec:classifer}}
In Stage II, we train an MLP using weighted cross-entropy to classify the representations from Stage I as bonafide or spoof. For a given input, the model weights from Stage I are frozen and a self-attention branch is used to compute the representation. Note that the two self-attention branches are identical due to weight sharing. 

\subsection{Data augmentations \label{sec:augs}}

We need a suitable data augmentation policy to support our training framework in three ways: prevent overfitting, handle speaker variability, and achieve robustness to telephony codec impairments. The following augmentations are used: pitch-shift, time-stretch, time and frequency masking from WavAugment \cite{wavaugment2020}, linear and non-linear convolutive noise and impulsive signal dependent additive noise from RawBoost \cite{Tak2021RawboostAR}, and the narrowband frequency impulse response (FIR) filters suggested in \cite{Tomilov2021STCAS}. All augmentations are applied on-the-fly during the model training.

%% file: 05.results.tex
\section{Experiments and results}
\label{sec:results}

\begin{table}[]
    \caption{EER performance of SSAST-CL on ASVSpoof 2021 LA evaluation set. Vanilla WCE is the policy recommended in \cite{gong2022ssast}. WCE-updated is the same as vanilla WCE, but with our suggested augmentation policy. Smaller EERs are better.}
  \centering
  \begin{adjustbox}{width=1\linewidth}
    \centering
\begin{tabular}{| l| l | c |}
\hline
\textbf{Training policy} & \textbf{ Augmentations} & \textbf{EER} \\ \hline
Vanilla WCE & \begin{tabular}[c]{@{}l@{}}Mixup, SpecAugment, \\ Noise\end{tabular} & 19.48 \\ \hline
WCE-updated & \begin{tabular}[c]{@{}l@{}}WavAugment, Pitch Shift, \\ Time Stretch, RawBoost,\\ Narrowband FIR\end{tabular} & 8.96 \\ \hline
\begin{tabular}[c]{@{}l@{}}\textbf{SSAST-CL (Ours)} \\ (self-attention only) \end{tabular}  & \begin{tabular}[c]{@{}l@{}}\textbf{WavAugment, Pitch Shift,} \\ \textbf{Time Stretch, RawBoost,} \\ \textbf{Narrowband FIR}\end{tabular} & \textbf{5.64} \\ \hline
\begin{tabular}[c]{@{}l@{}}\textbf{SSAST-CL (Ours)} \\ (self and cross-attention, $\alpha=0.2$) \end{tabular} & \begin{tabular}[c]{@{}l@{}}\textbf{WavAugment, Pitch Shift,} \\ \textbf{Time Stretch, RawBoost,} \\ \textbf{Narrowband FIR}\end{tabular} & \textbf{4.74} \\ \hline
\end{tabular}
    \end{adjustbox}    \label{tab:finetuning_baselines}
        \vspace{-10px}
\end{table}

\noindent\textbf{Audio pre-processing.} Raw audio waveforms of length $6$-seconds are used to create log-mel spectrograms of size $128$ mel-frequency bins and $512$ timebins, computed using PyTorch Kaldi~\cite{pytorchkaldi} with $25$ms Hanning window and $10$ms overlap. Longer waveforms are cut off at the end, whereas shorter waveforms are repeat padded by concatenating the original signal with its time-inverted version until the length is $6$ seconds.

\noindent \textbf{Sampling and batching.}
For the contrastive learning in Stage I, data pairs $(x_1, x_2)$ are created as follows: The sample $x_1$ is picked up in sequence from the training dataset, covering each datapoint once over a training epoch. For each $x_1$, we make use of its class information $c(x_1)$ to select the pairing sample $x_2$ such that positive and negative pairs are created with equal probability of $0.5$. When picking up $x_2$ from the spoof class, we assign equal probability of $0.5$ to the text-to-speech (TTS) and voice conversion (VC) subclasses. Once a pair is picked, each of the data augmentations from Section~\ref{sec:method} are applied to $x_{1}$ with a probability of $0.8$, and a randomly selected subset of data augmentations is applied to $x_{2}$. The above procedure is repeated until a batch of $64$ pairs is created. 
During Stage II, we pick batches of size $64$ sequentially from the dataset; to each sample, all data augmentations are applied  with $0.8$ probability.

\noindent\textbf{Training policy.} For both Stage I and II, we use Adam optimizer with a learning rate of $10^{-4}$, and an exponential rate decay scheduler with $\gamma=0.95$ for every $5$ epochs. The model is trained for $50$ epochs for each  stage. For Stage I, the epoch checkpoint reporting the least validation loss is chosen. For Stage II, the epoch checkpoint reporting the smallest validation EER is chosen.

\noindent \textbf{Implementation of SSAST-CL.} For details on the SSAST-CL implementation, please see the supplementary material. 

We now present numerical studies to demonstrate the improvements in EER when using our CL framework. We compare with the state of the art and ablate on the data augmentations.

\subsection{Impact of data augmentation, contrastive learning and cross-attention} \label{sec:contrastive_impact}

In Table \ref{tab:finetuning_baselines}, we compare the performance of the proposed contrastive learning framework against baselines that finetune the pretrained SSAST using weighted cross-entropy (WCE). The vanilla WCE, which follows the training and augmentation policy recommended in \cite{gong2022ssast}, reports a high EER of $19.48$ on ASVSpoof 2021 LA evaluation set. This is likely because the augmentations in vanilla WCE do not account for codec impairments. When our augmentation policy from the ablation studies in Table \ref{tab:data_augs} is used, the resulting WCE-updated policy reports an EER of $8.96$, which is marginally better than the $9.26$ reported by the best-performing  baseline B$03$ in ASVSpoof 2021~\cite{yamagishi2021asvspoof}.
When the training framework is \textit{additionally} replaced with our SSAST-CL framework, the EER improves significantly to $4.74$.

Table \ref{tab:finetuning_baselines} also demonstrates the impact of introducing cross-attention. We see that the EER improves from $5.64$ to $4.74$ when we set a weight of $\alpha=0.2$ to the cross-attention loss term $L_{CA}$. The t-SNE plots in Fig. \ref{fig:tsne} suggest that the proposed SSAST-CL framework better disentangles the bonafide and spoof classes when compared to the vanilla WCE baseline. There is also an improvement from introducing cross-attention; although the visual differences are subtle in this case due to the high reduction in dimensionality  from $192$ to $2$ when plotting the t-SNEs.  

 While the impact of SSAST-CL is evident from Table \ref{tab:finetuning_baselines} and Fig. \ref{fig:tsne}, we note that our framework is not fully optimized to report the best EERs. A grid search on the training hyperparameters, including the optimizer, learning rate, batch size, and cross-attention weight, will likely boost the EERs further. 

\begin{table}[ht]
\caption{EER comparison with SOTA on ASVSpoof21 LA evaluation set. $(^*)$~{estimate based on authors' description of model.}}
 \centering
  \begin{adjustbox}{width=\linewidth}
\begin{tabular}{|l|l|l|l|l|}
\hline
\textbf{Model} & \textbf{DA} & {\textbf{EER}} & \multicolumn{1}{c|}{\textbf{\begin{tabular}[ c]{@{}c@{}}Parameters \\ (million)  \end{tabular}}} & \textbf{GMACs}\\ \hline
\begin{tabular}[c]{@{}r@{}}Wave2Vec2.0+\\ AASiST~\cite{tak2022automatic} \end{tabular} & RawBoost (On-the-fly) & 0.82 & 317.84 & 1050 \\ \hline
LCNN-LSTM~\cite{Tomilov2021STCAS} & RS Mixup + FIR (On-the-fly) & 2.21 & 0.5$^*$ & 21.77$^*$\\ \hline
ResNet-LDE~\cite{Chen2021PindropLS} & \begin{tabular}[c]{@{}l@{}}RIR+Background noise + \\ Frequency Masking\end{tabular} & 3.68 & 11.5$^*$ & 100.74$^*$ \\ \hline
\textbf{SSAST-CL (Ours)} & \textbf{\begin{tabular}[c]{@{}l@{}}RawBoost + WavAugment + \\ Pitch Shift + Time Stretch + \\ Narrowband FIR (On-the-fly)\end{tabular}} & {4.74} & {5.99} & 126.75\\ \hline
RawNet2 \cite{Tak2021RawboostAR} & RawBoost (On-the-fly) & 5.31 & 25.43 & 59.45\\ \hline
ResNet-34 \cite{Cohen2021ASO} & \begin{tabular}[c]{@{}l@{}}Cellular, Telephony, and \\ VoIP Codecs (Offline)\end{tabular} & 5.18 & 21.5 & 623.89\\ \hline
ASVSpoof21 B03~\cite{yamagishi2021asvspoof}  & \begin{tabular}[c]{@{}l@{}}None\end{tabular} & 9.26 & 0.27 & 10.75 \\ \hline
\end{tabular}
   \end{adjustbox}
    \label{tab:sota_comparison}
    \vspace{-10px}
\end{table}

\subsection{Comparison with ASVSpoof21 top-performing models}
In Table \ref{tab:sota_comparison}, we compare the performance of the proposed SSAST-CL system against the top-performing single system models on the ASVSpoof 2021 LA evaluation dataset. Firstly, our system comprehensively outperforms the challenge's best-performing baseline B$03$ \cite{yamagishi2021asvspoof}. We see that an audio ViT, with appropriate training and data augmentations, can indeed achieve competitive performance on the audio spoof detection task. Note that a vanilla WCE finetuning on the SSAST model, as suggested in \cite{gong2022ssast}, results in worse EER than the challenge baseline (see Table \ref{tab:finetuning_baselines}). Secondly, when positioned against the best-performing models, our system reports comparable EERs while being significantly smaller in size than most of them. The LCNN-LSTM \cite{Tomilov2021STCAS} is the only {\em lightweight} system that reports a smaller EER than us. Lastly, our augmentations are much simpler than in the ResNet-LDE~\cite{Chen2021PindropLS} and ResNet-34~\cite{Cohen2021ASO} systems, as we do not use external noise or impulse response datasets for the augmentations. 

\begin{table}[ht]
   \caption{Impact of augmentations on the model performance for ASVSpoof 2021 evaluation data. All experiments additionally use time-stretch, pitch-shift, time masking, and frequency masking.}
     \begin{adjustbox}{width=1\linewidth}
\begin{tabular}{|l|c | c | l|}
\hline
Model                     & RawBoost & \begin{tabular}[c]{@{}l@{}}Narrowband FIR \end{tabular} & EER  \\ \hline
 &  \xmark        &   \xmark & 8.69 \\ \cline{2-4} 
SSAST-CL (Ours) & \xmark & \cmark & 6.88  \\ \cline{2-4} 
    & \cmark &   \xmark & 5.34 \\ \cline{2-4} 
&  \cmark &   \cmark  & 4.74 \\ \hline
\end{tabular}
    \end{adjustbox}
    \label{tab:data_augs}
    \vspace{-15px}
\end{table}

\subsection {Ablation on data augmentations} 
 In Table \ref{tab:data_augs}, we ablate over data augmentation combinations for the proposed SSAST-CL framework. We note that the RawBoost and FIR augmentations are both crucial for the model to perform well. These augmentations differ in design but both help capture the telephony codec artifacts, although to different extents. Telephony impairments are known to be present in the ASVSpoof 2021 LA dataset. The remaining augmentations in our policy, namely, time masking, frequency masking, pitch shift, and time stretch, are now commonplace in spoof detection modeling. These are essential because they account for speaker variability and prevent the model from overfitting \cite{wavaugment2020}.

%% file: 07.conclusion.tex
\section{Conclusion}
\label{sec:conclusion}

In this work, we answered the question ``Can we leverage ViTs for the audio spoof detection task?" We conducted investigations using SSAST-CL (our adaption of the SSAST model~\cite{gong2022ssast} for contrastive learning). A vanilla finetuning of the pretrained SSAST with cross-entropy loss gave sub-optimal performance. To learn more discriminative audio representations, the proposed SSAST-CL framework used Siamese training with a cross-attention branch and a novel contrastive loss formulation. An MLP was later used to classify the learned representations as bonafide or spoof. We empirically showed that the SSAST-CL framework successfully disentangles the bonafide and spoof classes, and it helps learn better classifiers for the task at hand. The introduction of cross-attention, along with suitable augmentations, has allowed our system to achieve competitive performance on the ASVSpoof 2021 challenge. 

Several research directions are enabled as a result of our work. A joint training of the two stages in our framework, using a multi-task loss formulation, could likely improve the model performance and also reduce the training time. Another direction would be to build an importance sampling/training policy where hard-to-learn samples (for example, of the voice conversion type) are prioritized. The proposed contrastive learning framework can also be extended to other downstream audio tasks where limited training data is available, such as emotion recognition~\cite{satt2017efficient} and language identification~\cite{cai2018novel}.

%% file: 09.supp.tex
\section{Supplementary material}
\label{sec:supplementary}
\noindent\textbf{Implementation details for SSAST-CL.}
In Stage I, the SSAST-CL framework employs Siamese training with the SSAST architecture \cite{gong2022ssast}, where the learnable weights are shared across three training branches: two for self-attention and one for cross-attention. In the self-attention branch, an input log-mel spectrogram $x$ is encoded into its attention-based representation vector  $\mathbf{r}_{SA}$ using a series of transformer and MLP blocks. As illustrated in Fig. 1, the log-mel spectrogram is split into a sequence of patches of size $128\times2$ created with a stride of $1$ time-bin. Each patch is then linearly projected into a $1\times 192$ vector, to which its learnable positional embedding vector (of the same size) is added. The resulting sequence of $511$ $(1\times192)$ patch embeddings is fed into a series of $N=12$ transformer blocks, each with $3$ attention heads. Within each transformer block, learnable weights $(\mathbf{W}_q, \mathbf{W}_k, \mathbf{W}_v)$ are assigned to the query, key, and value matrices for each attention head. These weights and other learnable parameters in the SSAST architecture are shared across the three branches for Siamese training. The output of the final transformer block, of size $(511 \times 192)$, is passed through a mean pooling layer to obtain the final representation $\mathbf{r}$ of size $1\times 192$. The same procedure is followed in the cross-attention branch, except that the input for the first skip connection is the sequence of 1-D patch embeddings from the $x_2$ branch.

For the purpose of contrastive loss calculations, all the self and cross attention representations $(\mathbf{r}_1^{SA}, \mathbf{r}_2^{SA}, \mathbf{r}_{12}^{CA})$ are further sent through a projection MLP comprising LayerNorm, linear upsampling to $512$, followed by LayerNorm and ReLU activation. For Stage II, we only consider the $x_1$ branch and pass the representations $\mathbf{r}_1^{SA}$, obtained from {\em before} the final projection MLP in Stage \rom{1}, through a MLP classifier comprising a linear layer to downsample from $192$ to $128$, followed by BatchNorm and ReLU, and a linear layer with softmax for binary classification.

To initialize the SSAST weights in Stage \rom{1}, we use the pretrained weights provided by the authors  in \cite{gong2022ssast}. Please see \cite{gong2022ssast} for details on how the SSAST is pretrained in a self-supervised manner using a patch masking and reconstruction strategy. 

\noindent \textbf{Computing Infrastructure.} We run experiments on a $1$x NVIDIA A100 GPU instance with $30$ vCPUs ($2.4$GHz AMD EPYC 7J13)  and $200$ GB memory. The average training time per epoch is $15$ minutes for Stage I and $10$ minutes for Stage II. A majority of the training time is spent on our (sequential) sampling and data augmentation procedure.

\bibliographystyle{IEEEtran}
\bibliography{main}